# Look at That Distractor: Dynamic Translation Gain under Low Perceptual Load in Virtual Reality

Ling-Long Zou[a], Qiang Tong[a,*], Er-Xia Luo[a], Sen-Zhe Xu[b], Song-Hai Zhang[c,d], Fang-Lue Zhang[d,e]

[a]*Beijing Information Science and Technology University, Beijing, China*
[b]*University of Science and Technology Beijing, Beijing, China*
[c]*BNRist, Beijing, China*
[d]*Tsinghua University, Beijing, China*
[e]*Victoria University of Wellington, New Zealand*

## Abstract

Redirected walking (RDW) utilizes gain adjustments within perceptual thresholds to allow natural navigation in large-scale virtual environments (VEs) within confined physical environments (PEs). Previous research has found that when users are distracted by some scene elements, they are less sensitive to gain values. However, the effects on detection thresholds have not been quantitatively measured. In this paper, we present a novel method that dynamically adjusts translation gain by leveraging visual distractors. We place distractors within the user's field of view and apply a larger translation gain when their attention is drawn to them. Because the magnitude of gain adjustment depends on the user's level of engagement with the distractors, the redirection process remains smooth and unobtrusive. To evaluate our method, we developed a task-oriented virtual environment for a user study (n = 26). Results show that introducing distractors in the virtual environment significantly raises users' translation gain thresholds. Furthermore, assessments using the Simulator Sickness Questionnaire (SSQ) and Igroup Presence Questionnaire (IPQ) indicate that the method maintains user comfort and acceptance, supporting its effectiveness for RDW systems.

---

*Corresponding author
*Email addresses:* zll749486310@163.com (Ling-Long Zou), tongq85@bistu.edu.cn (Qiang Tong), lex@bupt.edu.cn (Er-Xia Luo), senzhe@ustb.edu.cn (Sen-Zhe Xu), shz@tsinghua.edu.cn (Song-Hai Zhang), fanglue.zhang@vuw.ac.nz (Fang-Lue Zhang)

## 1. Introduction

Virtual Reality (VR) is a computer-generated technology that creates immersive three-dimensional environments, known as Virtual Environments (VEs), which are typically unconstrained by the limitations of the physical environment (PE). In VR applications such as The Room VR: A Dark Matter, users can explore large-scale VEs far exceeding the physical space they occupy, enabling immersive and engaging experiences. However, enabling unrestricted navigation in such expansive VEs remains a significant challenge. A range of locomotion techniques have been developed to address this issue, including teleportation [1], walking-in-place [2], and flying [3]. While these approaches provide basic movement capabilities, they either disrupt the sense of natural locomotion or require expensive hardware, potentially diminishing immersion and user engagement.

To overcome spatial constraints while preserving natural walking experiences, Razzaque et al. [4] proposed Redirected Walking (RDW), a technique that subtly alters users' real-world walking trajectories to enable natural navigation in a VE without leaving the bounds of the PE. However, RDW alone cannot fully prevent collisions with physical boundaries during unconstrained exploration. To further enhance safety, Reorientation Techniques (ROT) [5] reset users' direction in the PE without altering their VE-facing orientation, though often at the cost of reduced immersion. To mitigate this issue, Peck et al. [6] introduced Improved Redirection with Distractors (IRD), which uses visual distractors to redirect user attention, making subtle manipulations—such as increased rotation gain—less perceptible. Subsequent studies [7, 8, 9, 10] confirmed IRD's effectiveness in masking gain changes and raising detection thresholds. More recently, Ringsby [11] extended the use of distractors to translation gain but did not measure perceptual thresholds.

Building on these findings, we hypothesize that dynamically adjusting translation gain during user locomotion while their attention is drawn by distractors can effectively reduce perceptibility and expand the usable threshold range. Furthermore, distractors not only support gain manipulation but also enhance user engagement and exploration enjoyment. We developed a low-cognitive-load VR task environment containing only visual cues. A distractor was introduced during goal-directed walking to modulate user attention, and translation gain was dynamically adjusted. To ensure the generalizability of our findings, we selected a widely adopted consumer HMD. According to Valve's December 2024 SteamVR hardware survey, the Oculus Quest 2 HMD maintains a dominant market penetration of 34.21%, with sustained growth momentum. Accordingly, we implemented the experiment using the Oculus Quest 2. To accommodate hardware limitations (i.e., lack of eye tracking on Oculus Quest 2), we designed a head-direction-based method to quantify user attention, enabling smoother gain transitions.

To the best of our knowledge, this is the first study that:

- Proposes a head-direction-based method for quantifying attention and introduces a dynamic translation gain scheme to enhance exploration experience;

- Demonstrates that dynamic translation gain can be rendered less perceptible when user attention is occupied by distractors in low-load tasks, and empirically determines its perceptual threshold.

Our results show that distractors effectively mask translation gain changes, slightly expanding the detection threshold. SSQ [12] and IPQ analyses confirm that our method improves user comfort and immersion. Moreover, our approach is generalizable to low-cognitive-load VE with distractors, offering an effective solution for immersive VR locomotion.

## 2. Related Works

### 2.1. Redirected Walking and Distractors

RDW was proposed by Razzaque et al. [4] to enable natural navigation of vast VEs within limited PEs. This study pointed out that human self-motion perception primarily integrates vestibular, visual, and auditory cues. The inherent sensory noise in these systems makes humans particularly insensitive to additional rotation gain during head rotation. Preserving consistency among multiple perceptual cues increases the likelihood that humans attribute rotations to self-motion rather than to external environmental movements. Under this premise, RDW adjusts the user's virtual rotation angle, thereby redirecting the physical walking trajectory to enable virtual exploration.

RDW cannot completely prevent collisions with boundaries or obstacles. In this case, ROTs became essential to adjust user orientation within PE and ensure safe navigation. Some ROTs have now matured considerably. For instance, Li et al. [13] introduced a predictive method that estimates the probability distribution of users' positions and proactively plans the reset direction to make better use of the physical space. Building on the idea of optimizing spatial utilization through intelligent control, they [14] proposed a DRL-based framework that leverages spatial walkability entropy to guide users toward safer and more walkable areas, and enhances reset strategies by maximizing regional entropy. However, Peck et al. [15] indicated that ROT would compromise presence during virtual exploration. To mitigate this effect, they proposed integrating distractors into ROT, which helps maintain presence by redirecting user focus from the reorientation process. Subsequently, they introduced the IRD [6]. Specifically, they timed the distractor's appearance to coincide with when users were about to reach the PE boundary, with the distractor then moving back and forth in front of the user to capture attention. Following this, many studies related to distractors have emerged. Chen et al. [7] designed a dragon to interact with users to explore the practical application of distractors. Williams et al. [9] measured the effects of field of view (FOV), gender, and distractor on human perception of rotation gain. Kim et al. [16] showed from a different perspective that distractors can influence users' visual perception.

Ringsby [11] measured the detection threshold for rotation, curvature, and translation gains in the presence of a distractor. In his work, he established appropriate parameters for measuring rotation and curvature gains based on prior work [17], while adopting a conservative parameter set for translation gain. Regarding translation gain, no statistically significant effects were observed in the results, though all measured values remained within acceptable ranges for subjects. Furthermore, one study [18] employed a weighted combination of the user's gaze direction and previous locomotion direction to determine the translation gain adjustment direction. In this paper, we adapt the above design concepts by correlating the translation gain with the extent of user attention toward the distractor and utilizing a formula to ensure smooth gain variation.

*2.2. Translation Gain and Detection Threshold*

RDW research can be categorized into overt manipulation and subtle manipulation [19]. Overt manipulation serves to confine users' movements within the PE, employing techniques such as ROT and perceptible spatial transformations that violate real-world principles [20, 21]. This manipulation may interrupt continuous locomotion and reduce immersion. In contrast, subtle manipulation applies imperceptible adjustments during user locomotion, maximizing the duration before collisions occur to ensure a good experience.

Subtle manipulation employs gains to introduce imperceptible modifications to locomotion [22]. Specifically, gains represent the transformation of users' real motion data into VE coordinates through predefined mapping rules. Equation 1 illustrates the computation formula for translation gain:

$$g_t = \frac{T_{virtual}}{T_{physical}} \quad (1)$$

Where $T_{physical}$ denotes the physical locomotion distance, and $T_{virtual}$ represents the distance moved within the VE.

To preserve the natural walking experience, Steinicke et al. conducted a comprehensive discussion into optimal gain. Through a series of studies [23, 24, 25], they ultimately detected the translation gain threshold range as [0.86, 1.26] using two-alternative forced choice (2AFC). Subsequently, researchers extended this result by measuring translation gain threshold across various conditions. Kruse et al. [26] considered the holistic nature of visual stimuli and found that environmental richness outweighs visual self-representation in significance. Zhang et al. [27] projected 360° video into VE, establishing a translation gain threshold range of [0.942, 1.097], which indicates that higher fidelity enhances sensitivity to gain changes. Luo et al. [28] found that reducing the user's FOV through zooming can expand the threshold of translation gain. Moreover, Zhang et al. [29] found that gradual changes in dynamic translation gain permit significantly wider operational ranges while maintaining user imperceptibility.

In summary, translation gain thresholds are influenced by numerous factors. Contemporary researches primarily aim to expand thresholds while preserving natural locomotion perception, which is a focus that equally guides our study.

## 3. Preliminary Work

Before the experiment, we conducted preliminary work to aid in designing the experimental setup. Lavie et al. [30] used the inattentional blindness paradigm to examine load effects on conscious perception, in line with the predictions of load theory. Specifically, when users experience higher cognitive load (e.g., during multi-item search tasks that require target discrimination rather than simple identification), their awareness of task-irrelevant distractors is significantly decreased. Therefore, we needed to use low-load tasks to collect data more effectively. However, it was unclear whether the principles of load theory applied to VR tasks. To verify this and support the design of experiment, we constructed a VE to implement the aforementioned experimental procedure.

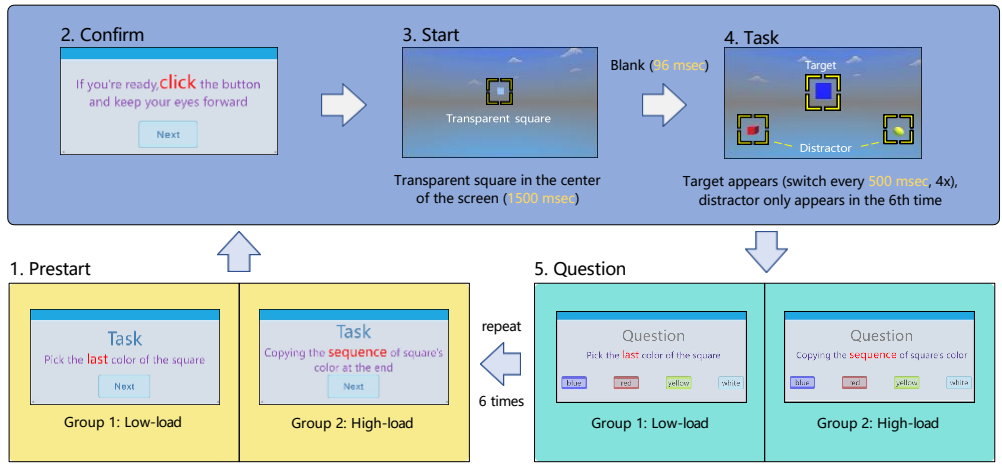

Figure 1: Preliminary work procedure.

The The experimental procedure is shown in Fig. 1. Each participant completed six trials. The first five trials contained only the target in the VE, while the final trial introduced an additional distractor during the *Task* phase. After answering the question, participants were asked if they had noticed any new elements that had not appeared in previous trials. They verbally described at least three relevant features (e.g., location, color, shape, and size relative to the target). Only participants who met this criterion were included in the analysis.

We employed the same data processing method as described in [30]. Participants with fewer than four correct responses were excluded from analysis 4 participants: 3 males and 1 female). Three excluded participants belonged to the high-load condition and one to the low-load condition. Subsequent analyses included data from the remaining 28 participants (14 per condi-tion).

Comparing the data from two load conditions, the unqualified rate for the high-load group (M=23.5%) exceeded that of the low-load group (M=12.2%). Significantly fewer participants detected distractors in the high-load condition (4 of 14) compared to the low-load condition (10 of 14), with a chi-square test yielding $\chi^2(1, N = 28) = 5.143, p < .05$. This result points to the fact that in VR, the level of perceptual load associated with relevant tasks affects awareness of irrelevant distractors. Furthermore, the distractor's placement in this scene matches our main experimental design, supporting the validity of the distractor positioning in VR.

## 4. Methods

### 4.1. Equipment

As the Oculus Quest 2 HMD currently dominates the consumer VR market, it is our device of choice for conducting experiments. It boasts a resolution of 1832 × 1920, a refresh rate of 90 Hz, and offers a FOV of 90 degrees under binocular overlap conditions. All experiments were performed on a computing system featuring an Intel Core i7-12700H processor (2.30 GHz), 16 GB RAM, and an NVIDIA GeForce RTX 3060 Laptop GPU. The VE was developed using Unity3D 2021.3.25f1c1 engine (with SteamVR plugin integration), while the experimental workflow was implemented through the OpenRDW framework [31]. During experimental sessions, we utilized Meta Quest Link software to establish wireless PC-VR streaming, enabling real-time rendering of VE on the HMD.

### 4.2. Dynamic Adjustment of Translation Gain

Given that the Oculus Quest 2 HMD lacks built-in eye-tracking capabilities, we designed a head direction-based method to quantitatively measure user attention as a proxy for gaze orientation. Prior research [33] has shown that, in immersive VEs, users tend to align task-relevant targets with the center of their visual field when performing tasks. This behavior supports the use of head pose as a reliable proxy for gaze direction in VR [34], especially for rough attention estimation. In that study, a revised logistic function was used to model object-based visual attention, applying a 15° gaze angle threshold, comprising 10° for central focus and an additional 5° as a tolerance margin. Inspired by this approach, we introduced an attention degree to dynamically modulate translation gain in relation to distractors. The attention degree is computed using the following equation:

$$A(t + \Delta t) = \begin{cases} A(t) + a \cdot e^{-\frac{deg^2}{2c^2}} \cdot \Delta t &, \ deg < d \\ A(t) - b \cdot \Delta t &, \ deg \geq d \end{cases} \quad (2)$$

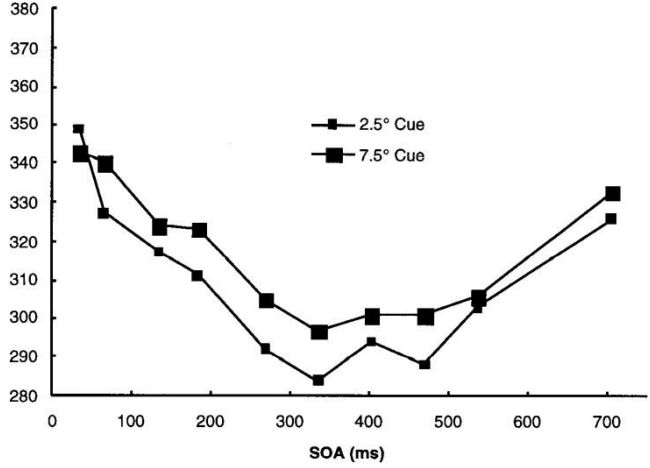

Figure 2: The curves describing the relation between Stimulus Onset Asynchrony (SOA) and reaction time in [32].

Where $A(t)$ represents the attention degree at time $t$, and $A(t+\Delta t)$ represents the attention degree at the next frame after time $t$, which is an accumulated value. Let $deg$ denote the gaze angle between the user's viewpoint and the centroid of the distractor, with the threshold parameter $d(=15)$ defining the angular detection boundary. The coefficients $a$, $b$, and $c$ represent hyperparameters, whose optimization yields a maximally smooth transition profile. In short, the attention degree we calculated is not the user's psychological attention, but rather closer to gaze alignment probability.

Benso et al. [32] pointed out that users' active focus duration is usually below 500 ms, and it fades out around 700 ms. Based on our observation of Fig. 2, we set the dynamic gain adjustment process to occur around 550 ms. We segmented the process into three phases based on the mean reaction time (MRT) trend profile: (1) ascent phase ($t_1$:200 ms), (2) maintenance phase ($t_2$: 300 ms), and (3) descent phase ($t_3$: 50 ms). The attentional focus adaptation period ranges between 33-66 ms [32]. Consequently, during the initial 33 ms of $t_1$, the attention degree remains constant. Subsequently, the gain gradually reaches its maximum as the degree of attention increases. Then the system transitions into $t_2$. Following 300 ms of sustained gain maintenance, phase $t_3$ initiates, wherein the gain linearly returns to 1.0 over 50 ms.

Based on the above design, we use a piecewise function to compute the attention degree. An exponential function is employed to model the increase in the attention degree, thereby reducing the initial sharp rise in gain. Furthermore, when the user stops gazing at the distractor, we use a linear function to steadily decrease the attention degree, avoiding abrupt changes in speed. According to users' feedback, we iteratively optimized *a*, ultimately determining that setting *a* to 5000 yielded the most natural adjustment dynamics. Parameter *b* was set to 2000, which corresponds to 50 ms time duration. *c* was set to 3.1 to ensure smoother transition dynamics at boundary values, making the increment approach zero as *deg* approaches *d*. In the implementation, we used the built-in Lerp method to perform continuous mapping between two frames.

Having established the hyper-parameter values, we proceeded to verify whether these settings could constrain the duration of phase $t_1$ within 200 ms. For this preliminary assessment, we employed a simplified version of the formal experiment and collected a total of 1100 measurements of $t_1$ duration from 20 participants. The recorded durations ranged from 0.060 to 0.713 seconds, with a median of 0.119 seconds, a mean of 0.123 seconds, and a standard deviation of 0.044 seconds. These results suggest that the selected hyper-parameters not only preserved the naturalness of the process but also effectively maintained $t_1$ within the desired 200 ms threshold.

## 5. Experiment

This experiment aimed to validate our method can make gain variations relatively less perceptible while simultaneously improving users' experience.

### 5.1. Experiment Design

To evaluate the impact of distractors and gain-switching strategies, we established one experimental group and two control groups. In the first control group, participants performed the task without the presence of any distractors. In the second control group, translation gains were directly switched based on predefined values, without being influenced by interaction dynamics. All experiments were conducted in a 10m × 10m obstacle-free physical environment. To promote immersion and streamline data collection, we designed a low-cognitive-load virtual scenario, as illustrated in Fig. 3. In this VE, participants were tasked with escorting a virtual cat to a target location by maintaining a fixed distance and deterring potential threats—modeled as virtual humans or dogs—that emerged along the path (Fig. 4). The spatial layout was designed to ensure task consistency and spatial feasibility. The participant's starting point was marked by a blue circle, the cat's initial location by a white circle, and the destination by a red circle. Distractors were placed on the lateral edges of the scene, initially hidden behind houses. The cat always traveled a fixed distance of 8m to its destination, guaranteeing that no matter the gain setting, the participant's path remained within the bounds of the physical space. A trigger zone, located 1.5m ahead of the cat, activated the appearance of a distractor, which would then move at a constant speed toward the path center. Each distractor had a hidden progress bar reflecting the level of visual deterrence achieved through participant gaze. Upon reaching the maximum deterrence threshold, the distractor changed its behavior and the translation gain increased to its maximum value. In addition, we use a Boolean field to record whether the data has reached the maximum gain. Data that do not meet the criteria are excluded from the subsequent analysis. Given prior findings that realism in virtual environments does not significantly affect user immersion [27], we opted for a low-poly, cartoon-style design to reduce scene complexity.

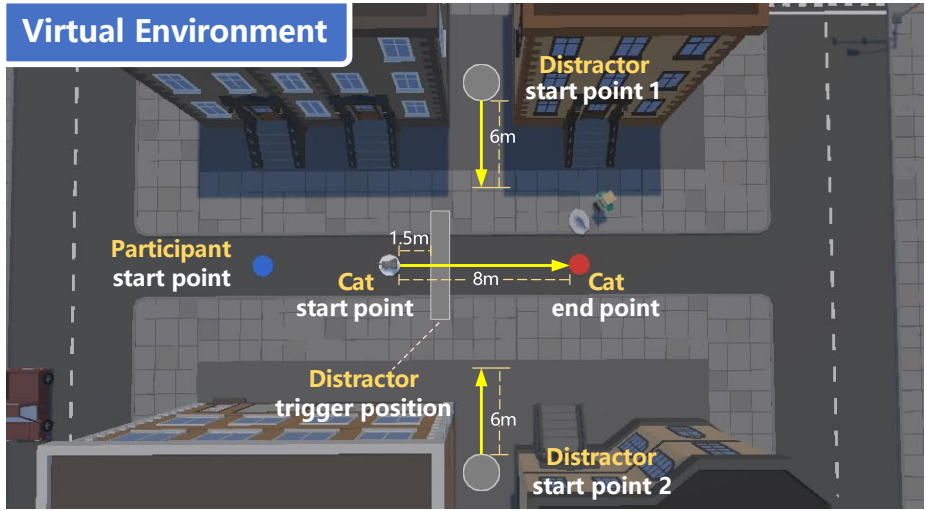

Figure 3: Virtual environment. Important points are marked in the figure. For the participants, we mark only the starting point because trial completion depends on whether the cat reaches the end point. Distractor starting points are visually blocked by buildings to prevent participants from predicting their locations.

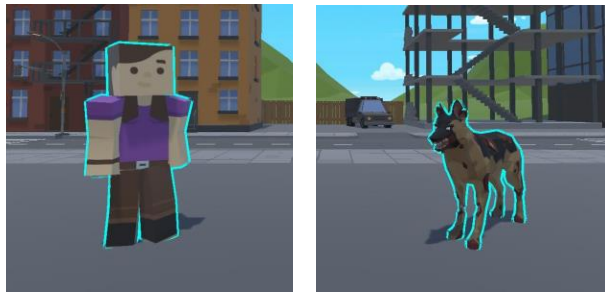

Figure 4: Distractors in experiment. The human distractor (left) and the dog distractor (right) are used because they fit well in our scene. Humans and animals can serve as distractors in most virtual scenes, as they are common elements that naturally blend into VEs.

We conducted a comprehensive user study based on a 3 (Group: with distractors, without distractors, gain switch) × 11 (Gains: 0.5 to 1.5 in 0.1 increments) design. Each gain setting was tested five times per group, resulting in 165 trials per participant. Before the experiment began, we randomized the order of gain presentations using the Fisher-Yates shuffle algorithm, generating five unique sequences stored in a configuration file. At the start of each group block, the system loaded a predefined gain sequence from the file. This setup ensured that while the gain order varied between trials within each group, the sequence remained consistent across all three groups for each participant—enabling controlled comparisons across conditions.

Overall procedure of experiment is illustrated in Fig. 5. In the 'w/ dst' group, distractor would emerge from randomized lateral positions when the cat reached trigger zone. In the 'w/o dst' group, there was no more distractor during trials, with gain variation was controlled by predetermined time intervals and linear velocity profiles. Distinction of the 'switch' group from the 'w/o dst' group lies in the way of gain variation. The specific gain

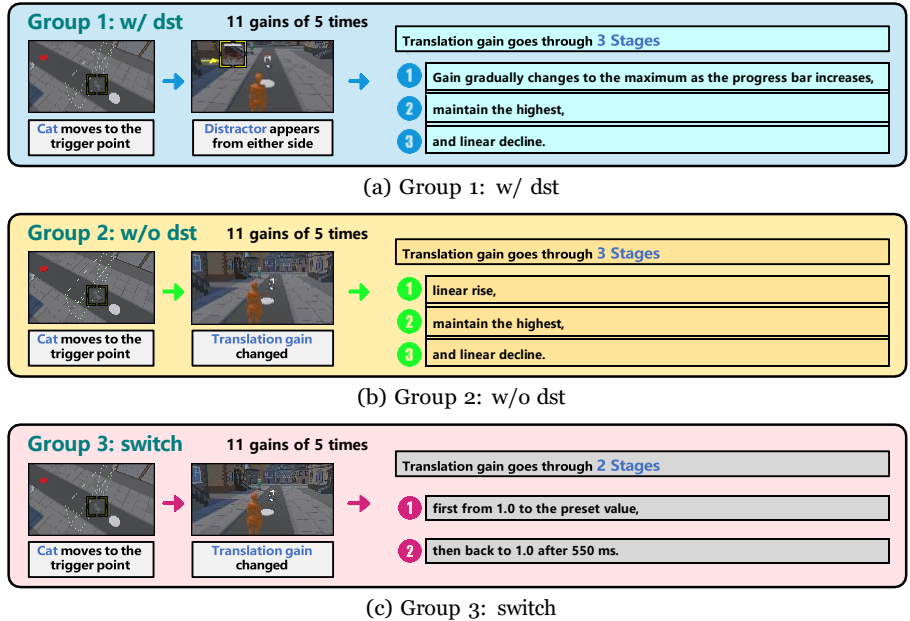

(a) Group 1: w/ dst

(b) Group 2: w/o dst

(c) Group 3: switch

Figure 5: Experiment procedure. The 11 gain values start at 1 and transition to their target values upon triggering.

switch patterns for each group are shown in Fig. 5.

After the trial ended, a question prompt appeared in front of the participants. The question was set as "Was your movement in the virtual world greater or smaller than in the physical environment". The total duration of the experiment was approximately 45 min.

*5.2. Participants*

26 participants, age 22 - 25 (13 females(M = 23.85, SD = 0.59) and 13 males(M = 23.77, SD = 0.64)) participated in experiment. All participants were graduate students with normal or corrected-to-normal vision, and without history of epilepsy or a strong predisposition to motion sickness. One of the females was left-handed and we provided her with a left-handed handle for the interaction. Of these, 15 participants had previous experience with VR, while the remaining 11 participants were novices to it.

*5.3. Procedure*

Before the experiment, all participants completed a preliminary trial. Participants were asked to experience their normal walking speed in this trial, which would serve as a reference for all subsequent trials. Before each trial, participants were instructed to stand on the blue marker and adjust their initial pose. Afterwards, they used the handle to press the button, initiating the trial. As participants walked, the cat moved proportionally the same distance. When the cat entered the trigger zone, corresponding events occurred in the scene. As the cat contacted the red circle, a question prompt appeared in front of the participants. After the participant answered the question, the system computed the real locomotion distance and obscure the visual display to reset the pose. This design lets participants move to a designated location after each reset, then rotate 180° for the next trial, avoiding a return to the starting point. Once they completed each group block, participants were asked to fill out the SSQ and IPQ to quantitatively assess their experience. After all trials were completed, we spoke with each participant.

The experiment and procedure were approved by XXX University Science and Technology Ethics Committee.

### 5.4. Results

#### 5.4.1. Detection thresholds

We recorded the user's choices and utilized the *quickpsy* library [35] in R to fit psychometric curves, with the results shown in Fig. 6. Our experimental tasks are categorical tasks with binary response variables and stimulus level serving as explanatory variables. These types of tasks are commonly modeled using a dichotomous approach:

$$f(k;\vartheta) = \prod_{i=1}^{M} \binom{n_i}{k_i} \psi(x_i;\vartheta)^{k_i}(1-\psi(x_i;\vartheta))^{n_i-k_i} \tag{3}$$

$f$ represents the probability mass distribution function for the model. $M$ represents the number of stimulus levels used in these types of tasks. $x_i$ represents the i-th stimuls level. $n_i$ is the number of times that $x_i$ is presented. $k_i$ represents the number of *Yes–type* (or correct) responses. $\psi(x_i;\vartheta)$ called the psychometric function and has the form as follow:

$$\psi(x;\vartheta) = \psi(x;\alpha,\beta,\gamma,\lambda) = \gamma + (1-\gamma-\lambda)F(x;\alpha,\beta) \tag{4}$$

$\alpha$ and $\beta$ are the position and scale parameters. $\gamma$ and $\lambda$ are the parameters corresponding to the leftward and rightward asymptote of $\psi$. $F$ is a function

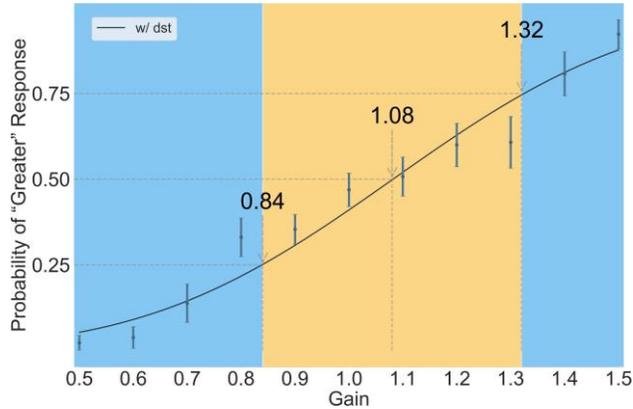
(a) Group 1: w/ dst

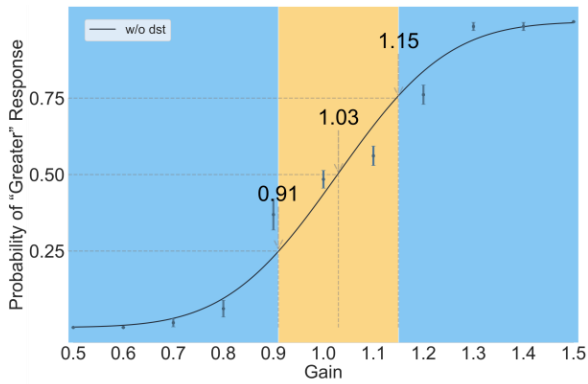
(b) Group 2: w/o dst

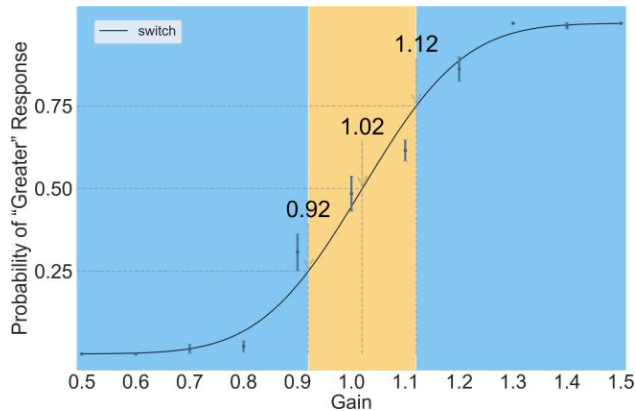
(c) Group 3: switch

Figure 6: Gain detection threshold. Y-axis represents the probability of "Greater" response, and X-axis represents the gain value. Three dashed arrows in each group mark the lower detection threshold (LDT), point of subjective equality (PSE), and upper detection threshold (UDT).

with leftward asymptote 0 and rightward asymptote 1. In the *quickpsy* library for R, use the following cumulative normal function as *F*:

$$F(x; \alpha, \beta) = \frac{\beta}{\sqrt{2\pi}} \int_{-\infty}^{x} \exp\left(\frac{-\beta^2(x-\alpha)^2}{2}\right) \quad (5)$$

The fitted curves model participants' responses to different levels of stimulation. The gain at which 50% of participants selected "Greater" corresponds to the PSE. Threshold detection aims to determine the LDT and UDT. The LDT and UDT are defined as the gain values where participants respond "Greater" with probabilities of 25% and 75%, respectively.

All fitted models exhibited S-shaped curves, with AIC values of [88.2, 73.2, 59.6], and SSE values of [0.0397, 0.0404, 0.0228], which indicate acceptable model fits. The thresholds were [0.84, 1.32] for the 'w/ dst' group, [0.91, 1.15] for the 'w/o dst' group, and [0.92, 1.12] for the 'switch' group. Moreover, the 95% confidence intervals for the PSEs were [1.07, 1.10], [1.01, 1.04], and [1.00, 1.03] for the three groups, respectively. From the results, it can be observed that the PSEs for all three groups were slightly greater than 1.0, indicating that the majority of participants tended to underestimate travel distances in the VE. To validate the efficacy of the experimental group method, we further analyzed the raw data and the model-fitted data.

We derived the probability of responding "Greater" from the raw data and compared these with a 3 (Group: w/ dst, w/o dst, switch) × 11 (Gain: 0.5 to 1.5 in 0.1 increments) × 2 (Gender: female, male) ANOVA. The factor of Gain showed a significant effect ($F = 488.25, p < .001, \eta^2 = .54$), and there was also a significant effect of Group ($F = 7.548, p < .01, \eta^2 = .04$). Additionally, a significant Group × Gain interaction ($F = 12.703, p < .001, \eta^2 = .06$) was observed, while all other interactions were non-significant. The factor Gender had no significant effect ($F = .034, p = .587, \eta^2 = .01$), and combining this with the above analysis, it suggests that gender has little impact on the results.

Regarding the fitted curve data, we preformed a 3 (Group: w/ dst, w/o dst, switch) × 2 (Gender: female, male) ANOVA. There was a significant effect of Group ($F = 2.978, p < .05, \eta^2 = .03$). Post-hoc analysis of pairwise comparisons was performed with Bonferroni adjustments. There were significant differences between the 'w/ dst' group and the other two groups (all with $p < .05$), which suggests that the effect of using gain dynamic adjustment method alone is weak.

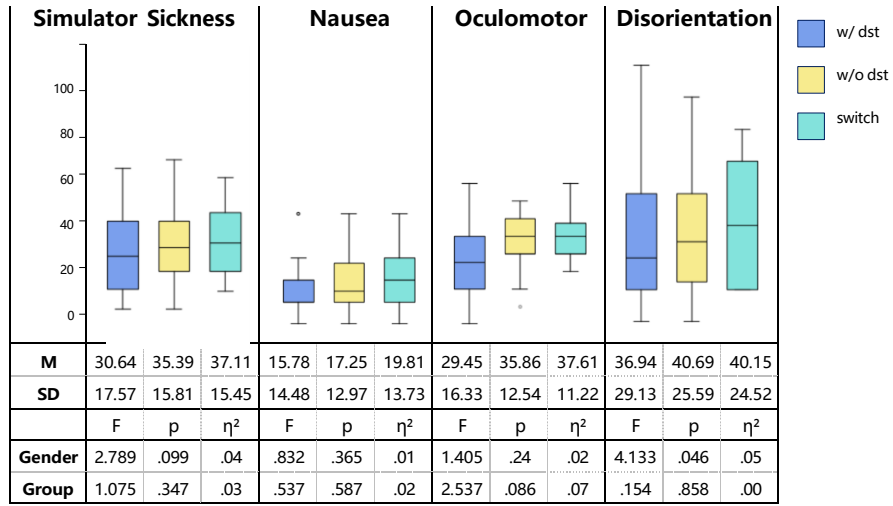

Figure 7: SSQ scores. The leftmost represents the total severity (TS) score, and the three items to the right correspond to the three dimensions of simulator sickness.

*5.4.2. Simulator Sickness*

The SSQ comprises 16 simulator-related symptoms, each rated on a 4-point severity scale, specifically developed to quantify VR-induced sickness. The detailed SSQ scores for the experiment are shown in Fig. 7. We conducted a 3 (Group: w/ dst, w/o dst, switch) × 2 (Gender: female, male) ANOVA on TS scores and found no significant effects for any of the variables (Group: $F = 2.789, p = .099, \eta^2 = .04$, Gender: $F = 1.075, p = .347, \eta^2 = .03$). Furthermore, we conducted separate analyses of the three SSQ subscale scores, none of which reached statistical significance (as shown in Fig. 7).

Table 1: IPQ scores.

| Group | G | SP | INV | REAL |
| --- | --- | --- | --- | --- |
| Group 1: w/ dst | 5.15 | 4.79 | 3.65 | 4.13 |
| Group 2: w/o dst | 5.04 | 4.45 | 3.21 | 3.66 |
| Group 3: switch | 4.81 | 4.17 | 3.16 | 3.19 |

*5.4.3. Sense of Presence*

The IPQ encompasses four subscales: Spatial Presence (SP), Involvement (INV), Experienced Realism (REAL) and Overall Sense of Presence (PRES). PRES is obtained by averaging the mean scores of the other three subscales[36]. This 14-item questionnaire is specifically designed to quantify users' presence perception in VE. We present the IPQ scores in Table. 1. We conducted a 3 (Group: w/ dst, w/o dst, switch) × 2 (Gender: female, male) ANOVA on IPQ data and consistently observed significant effects of Group across all three subscales (SP: $F = 5.967, p < .05, \eta^2 = .08$, INV: $F = 3.466, p < .05, \eta^2 = .09$, REAL: $F = 20.858, p < .001, \eta^2 = .37$). Post-hoc analysis of pairwise comparisons was performed with Bonferroni adjustments. There were significant differences between the 'w/ dst' group and the other two groups (all with $p < .05$). However, no significant effects were detected for Gender and the interaction between the two.

## 6. Discussion

*6.1. Detection Thresholds*

During the initial design phase, the 'w/o dst' group was designed to verify the effect of distractors, while the second control group aimed to validate the effectiveness of the dynamic translation gain variation method. However, the results reveal no significant difference between the two, suggesting that the method alone has only a minimal effect. The threshold for the experimental group was [0.84, 1.32], broader than that reported by Steinicke [25]. However, since our measured translation gain differs from the conventional definition, making direct comparisons invalid. Nevertheless, significant effects were observed between the experimental group and both control groups. This provides evidence against the hypothesis and demonstrates that our method, when combined with distractors, can effectively expand the threshold range.

In addition, we conducted analysis on verbal communication after the experiment. Some participants felt that they could distinctly detect changes in their own speed, especially when looking straight ahead. As one participant explained, "It felt like being suddenly pushed forward by something, the optical flow in my peripheral vision sped up". When distractors were present, participants focused primarily on the distractors' movements, resulting in a higher probability of overlooking gain changes.

In fact, we believe this does not represent the full potential of our method. Since the translation gain changed over time, participants based their judg-ments on the initial gain. It is only when there is a significant difference between the initial and maximum gains that participants can perceive inconsistencies in their walking speed. At the gain of 0.9 or 1.1, participants' perception of walking speed changes was highly ambiguous, with most reporting inability to discern speed. If we select different initial gains, the threshold may expand even further. Nevertheless, this hypothesis requires further experimental validation.

## 6.2. User Experience
### 6.2.1. Simulator Sickness

Unlike the control group, participants in the experimental group had to turn their heads laterally to look the distractors. Based on prior knowledge, this head-turning behavior could potentially exacerbate motion sickness symptoms. Quantitative results showed no significant differences between the groups. However, the experimental group exhibited greater standard deviation, suggesting that some participants were still affected. Initially, we hypothesized that gender differences might account for this observation. Nevertheless, statistical analysis revealed no significant gender effect. After excluding VR motion-sensitive factors, our method did not show any significant exacerbation of motion sickness severity during use.

### 6.2.2. Sense of Presence

The IPQ results indicated that the experimental group scored higher across all subscales compared to both control groups. To further validate the results, we conducted significance analyses that revealing the factor of Group's significant effects on the three subscales. Additionally, no significant gender effects were observed in the results. Through verbal feedback, participants indicated that the presence of distractors enhanced their interactive experience during the tasks, making them more willing to immerse themselves in the VE. Combined with the results of quantitative analysis, this effectively demonstrates that our method can enhance the user's exploration experience.

## 7. Limitations and Future Works

### 7.1. Limitations
#### 7.1.1. Initial Gain

A limitation of our study is that it only investigated scenarios where the translation gain changed from 1.0, without checking cases with other initial values. To explore the cases with different initial values, the future work can start with selecting appropriate upper and lower boundary values as the initial values. For instance, a work [29] chose 0.7 and 2.0 as two additional initial values. However, our preliminary tests showed that a constant translation gain of 0.7 is perceived as noticeably slower, whereas 2.0 is perceived as an abnormally high speed. Therefore, we did not adopt the selection of gains from this study. In contrast, we hope to explore a more general research methodology that can determine the threshold conditions for dynamic changes in all gain values within the range of [0.86, 1.26]. Worth noting, we observed that when the variation between altered gain and original gain is small, participants are highly unlikely to detect the gain change. This would allow for a broader of possibilities in the use of dynamic translation gain. However, we currently lack an effective solution to implement this idea and aim to address it in the future work.

*7.1.2. Diversity of the Participants*

Due to time constraints in recruiting participants for the experiment, we chose to recruit only from within our university. This led to a participant pool composed exclusively of graduate students aged 22 – 25, indicating a lack of demographic diversity and potentially limiting the generalizability of our findings to broader user populations.

*7.1.3. Control Group*

After the experiment, we believe a control condition with a constant gain should be introduced to isolate the effect of the distractor. This could help us better understand the interaction between external distractions and dynamic gain modulation.

*7.1.4. Learning Effect*

A limitation of our study is the lack of randomization in the timing of gain changes across groups, which may have introduced learning effect. Ideally, all groups would use randomized gain-change timings to eliminate this confound. We had tried triggering the distractor from different positions, but this resulted in a significant increase in the distance users needed to walk. In the end, we were unable to find a satisfactory compromise solution.

*7.1.5. Selection of Gain Change Time*

The 550ms duration set in this paper is based on previous work [32], without being confirmed through additional experiments, so there might be a more suitable timing setting for the current scenario.

*7.2. Future Works*

*7.2.1. More Types of Gains*

We only inspected the dynamic changes of translation gain, whereas in most redirected controller work today, the three fundamental gains are used in conjunction. The other two gains can also be dynamically adjusted when the user's attention is diverted, which is theoretically feasible. With further refinement, this approach could be extended to paths with curves and turns. The current measurement setup is suitable for curvature gain, requiring only an appropriate physical space. For rotation gain, however, we would need to redesign the scenario and would refer to the design proposed by Williams et al. [9]. In the future, we will continue to explore the rotation gain and curvature gain, which may lead to the development of a more universally applicable RDW solution.

*7.2.2. Eye-Tracking*

Currently, we are researching a generalizable method that is better suited to the majority of consumers. In fact, in the future, we can adopt HMD equipped with eye-tracking to achieve more precise perceptual measurements. This will enable our research to evolve from being generalizable to being targeted, meeting higher-level usage requirements. Since eye-tracking data can better reflect users' attention, the adaptation time required during the initial phase of gain changes will be significantly reduced or even eliminated. At that point, we will need to redesign the relevant equations to accommodate it.

## 8. Conclusion

In this paper, we proposed a head-direction-based method that utilizes distractors to dynamically vary translation gain. By approximately quantifying user attention and using it as guidance to alter the translation gain, we make the change process smoother.

We first conducted a preliminary work based on perceptual load theory to investigate participants' detection of distractors under varying levels of perceptual load in VR. The results show that perceptual load theory is valid in VR. Furthermore, participants are more likely to notice irrelevant distractors under low perceptual load than under high perceptual load. Based on this finding, we designed the experiment using a low-load task.

In the experiment, we applied the dynamic translation gain adjustment method proposed in this paper. The results show that our method can slightly reduce users' perceived sensitivity in the presence of distractors. Furthermore, the results from the SSQ and IPQ demonstrate that our approach can enhance both user comfort and spatial presence across a broader range of thresholds, making it an effective RDW tool.

*8.0.1.*

This work was supported by the National Natural Science Foundation of China (62402281, 62361146854), the Fundamental Research Funds for the Central Universities (FRF-TP-25-036), the Young Backbone Teacher Support Plan of Beijing Information Science & Technology University (YBT 202426), and the Tsinghua-Tencent Joint Laboratory for Internet Innovation Technology.